# Cryostat for Ultra-low-energy Threshold Germanium Spectrometers

Craig E. Aalseth, Ricco M. Bonicalzi, James E. Fast, Todd W. Hossbach, John L. Orrell, Cory T. Overman, Brent A. Vandevender, *Pacific Northwest National Laboratory*

*Abstract*—This paper presents progress on the development of a cryostat intended to improve upon the low-energy threshold (below 0.5 keV) of p-type point contact germanium gamma-ray spectrometers. Ultra-low energy thresholds are important in the detection of low-energy nuclear recoils, an event class relevant to both dark matter direct detection and measurement of coherent neutrino-nucleus scattering. The cryostat design, including a thermal and electrical-field model, is given. A prototype cryostat has been assembled and data acquired to evaluate its vacuum and thermal performance.

*Index Terms*—cryostat, electrical-field model, low-energy threshold, p-type point contact germanium detector, thermal model.

## I. Introduction

RECENT reports have shown the research value of large volume (>100 cm$^3$), low-energy threshold (<0.5 keV) high-purity germanium (HPGe) gamma-ray spectrometers for dark matter searches [1] and coherent neutrino-nucleus scattering measurements [2]. Both applications seek to quantify the ionization generated by the recoil of a germanium nucleus struck by the weakly interacting particle of interest, i.e., a WIMP dark-matter particle or a neutrino. Due to the nature of such interactions, progressively more events occur having lower-energy nuclear recoils. Therefore, decreasing the detector energy threshold is a main approach to increasing detector sensitivity. State-of-the-art x-ray detector design suggests an energy threshold as low as 0.1 keV may be possible [3,4,5]. Avenues for pursuing lower-energy threshold HPGe detectors include modification of detector electrical contacts, detector mounting, and novel low-noise preamplifier design.

To provide a research and development platform for future investigations related to the desired ultra-low-energy threshold HPGe detector, a cryostat design was developed having three principal requirements: excellent thermal performance, amenability to fabrication from ultra-high purity, electroformed copper [6,7], and adaptable crystal mounting.

Thermal performance is important, as one component of noise from HPGe detectors is due to bulk leakage current from the detector. Although this current is primarily a characteristic of the specific germanium crystal, it is also true that higher operating temperatures result in higher bulk leakage currents. The second design requirement, though not utilized for the cryostat presented in this article, provides a way to produce an ultra-low background detector. Perhaps the most important design requirement is the capability to attach modified versions of the crystal mount to the cold finger. This capability facilitates studying the impact of mount design on the noise performance of the detector (which determines energy threshold). Furthermore, for versatility a commercial-style joint has been used between the cold finger and mount to allow interchanging the crystal mount between this cryostat and a commercial cryostat.

In this article is given the engineering design of the cryostat, thermal and electrical field models, initial performance results, and other studies relevant to choices made in the cryostat fabrication.

## II. Cryostat Design

A p-type point contact (PPC) HPGe detector [2] is the crystal configuration chosen for the focus of this effort to produce an ultra-low-energy threshold germanium gamma-ray spectrometer. The point-contact electrode structure produces a low detector capacitance (~1 pF), resulting in low electronic noise (and thus a low energy threshold). PPC detectors' low-noise performance, in combination with their large volume (>100 cm$^3$), makes these detectors unique in their sensitivity to low-energy nuclear recoils induced by rare weak-interaction events.

The cryostat design given attempts to take full advantage of the low detector capacitance. First and foremost, the crystal mount employs an "open" design where the crystal is held by small dielectric stand-offs around its edges as opposed to being fully enclosed within a dielectric structure, e.g., wrapped in PTFE plastic. This design reduces the stray capacitance due to dielectric material, reducing possible contributions to the electronic noise, which will negatively impact the energy threshold..

Manuscript received June 13, 2012. The Ultra-Sensitive Nuclear Measurement (USNM) Initiative, a Laboratory Directed Research and Development (LDRD) program at the Pacific Northwest National Laboratory supported this work.

All authors are with Pacific Northwest National Laboratory, Richland, WA 99352 USA. Corresponding author is R. M. Bonicalzi (phone: 509-375-5936; fax: 509-371-7869; e-mail: ricco.bonicalzi@pnnl.gov).

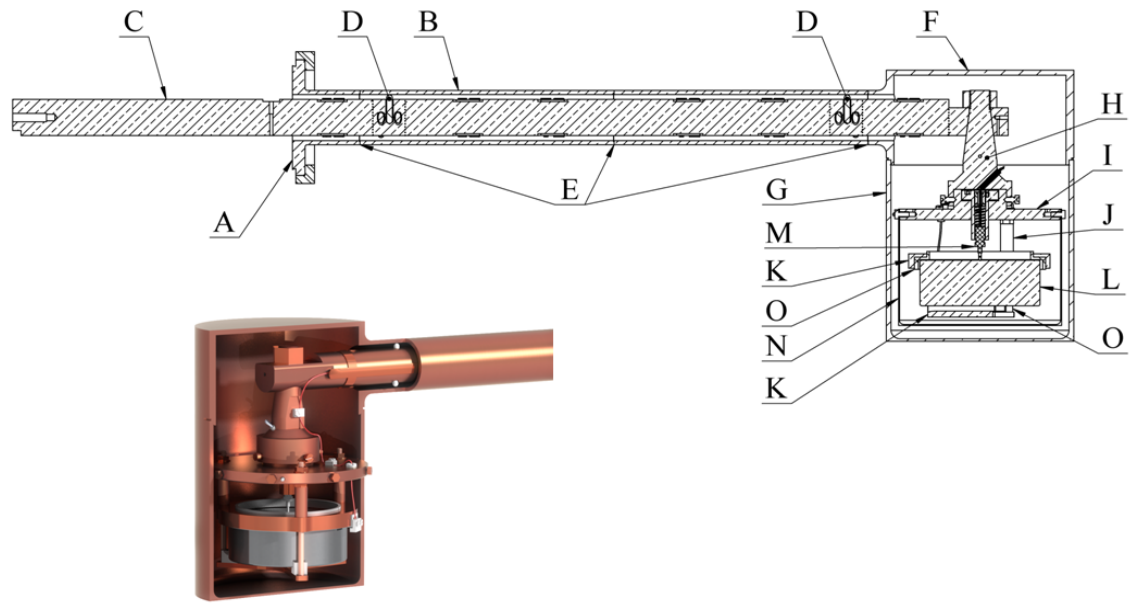

Fig. 1. Cryostat rendering and schematic including (A) dipstick flange, (B) crossarm, (C) horizontal cold finger, (D) PEEK stand-offs, (E) e-beam welds, (F) end cap, (G) end can, (H) vertical cold finger, (I) cold plate, (J) copper stand-offs, (K) copper backer plates, (L) germanium crystal, (M) signal contact, (N) infrared shield, (O) PTFE stand-offs.

Since this cryostat design is targeted for rare event detection, the cryostat is constructed using low-background methods, including use of radiopure materials and cleanroom assembly. Furthermore, the design is compatible with future fabrication of a majority of its elements from ultra-high purity, electroformed copper, for ultra-low-background applications. Fig. 1 shows the cryostat design with a generic HPGe PPC detector of approximately 1-kg mass

*A. Vacuum Jacket Design*

The vacuum jacket consists of a dipstick flange (A), crossarm (B), end cap (F), and end can (G), as shown in Fig. 1. The cryostat is designed to operate continuously for multiple years and will ideally have a minimal number of vacuum cycles and molecular-sieve regenerations. For this reason, minimizing the permeation of atmospheric gases into the cryostat suggests the choice of permanent joints over demountable or gasketed joints throughout the vacuum envelope, except where such joints are absolutely necessary. Permanent joints are made with an electron-beam welder and billet-machining practices are used to reduce leakage. E-beam welding is utilized since it requires no filler material and therefore avoids possible backgrounds due to radioactive impurities in the filler.

The joint between the cryostat end can (G) and end cap (F) is not permanent as removal of the can is necessary for changing the crystal mount. During testing this joint is sealed with an elastomeric o-ring, though for final assembly, high-purity indium wire will be used as the gasket material. Indium has a low permeation rate, low flow-stress values at room temperature, and cold-weld properties. The indium will tend to flow into any surface defects caused by machining or handling. A screwdriver slot exists for breaking the indium seal. An EGS4 [8] simulation was used to check that the radioactivity of the indium would not induce significant backgrounds in the detector. Because this system makes use of a commercial dipstick, a second impermanent joint in the form of an elastomeric o-ring exists between the sealing surfaces of the dipstick and cryostat body.

There are three e-beam joints (E) along the crossarm and each is configured for radial (as opposed to concentric) welding. This configuration simplifies the welding setup as the electron beam remains stationary while the part rotates about its cylindrical axis. Each mating part has a concentric locating surface so that only compression is needed to hold the joint together during the welding.

Precautions were taken to minimize virtual leaks internal to the cryostat. Whenever possible vented screws were used and blind-tapped holes and o-ring channels were vented. E-beam joints were configured to minimize trapped volume.

The vacuum pressure is maintained by molecular sieve in the bottom of the commercial dipstick which cryopumps residual gases resulting from joint-permeation, virtual leaks, outgassing of cryostat surfaces, and gases left over from turbo pumping. Experience suggests that after the first ~year of use, pumping on the cryostat and regenerating the sieve material to remove the gases from virtual leaks and outgassing is beneficial to long-term vacuum quality. Similarly, prior to mounting a germanium crystal in the cryostat, regeneration of the molecular sieve is advisable in order to maximize the cryopump capacity for long-term use. Regeneration also ensures that there is a net flux of gas moving away from the crystal surfaces.

Standard clean vacuum protocol is followed, e.g., nitrogen is used as a backfill gas to minimize water vapor in the cryostat. Since outgassing from dielectrics used in the cryostat is a concern for long-term vacuum quality, all dielectrics are pre-treated with heating and vacuum pumping in addition to chemical leaching after fabrication. Dielectrics are stored in a dry nitrogen environment while awaiting use in the cryostat assembly.

## B. Thermal Design

One factor affecting the energy threshold of a semiconductor detector is the leakage current of charge carriers thermally induced across the band-gap, contributing to the parallel white noise of the detector system [9]. HPGe detectors are typically operated near liquid-nitrogen temperature (77.2 K) though operating at temperatures as high as 130 K is feasible [10]. However, since the low-energy threshold goal for this cryostat requires minimization of all noise sources, a crystal temperature near that of liquid nitrogen is desired. Excellent thermal performance is important as leakage current increases exponentially with temperature, increasing by as much as a decade for a few K.

The thermal system consists of the vertical cold finger within the dipstick, horizontal cold finger (C), vertical cold finger (H), cold plate (I), and infrared shield (N), as indicated in Fig. 1. All these parts are made of OFHC copper. PEEK stand-offs (D) between the horizontal cold finger and crossarm support the weight of the cold finger and crystal mount. Stand-offs of PCTFE placed around the circumference of the cold plate prevent thermal shorts between the infrared shield and end can.

As part of the thermal design process, a simple thermal model estimates the temperature of the crystal will be 103 K. The values needed for this calculation are the thermal resistance of the path from the crystal to the liquid-nitrogen bath and the various heat loads on the system. The calculations are conservative in the values used for the joint thermal resistances and in treating all heat loads as if they are applied at the crystal.

The thermal model includes estimates for the electrical (0.07 W), conductive (0.5 W), and radiative (2.2 W) heat loads, which includes joule heating from the front-end electronics, conduction through the thermal stand-offs, high-voltage wiring and signal wiring, and thermal radiation from the vacuum jacket. The radiative-load calculation uses $\varepsilon=0.03$ for the room-temperature (300 K) copper emissivity [11,12], and based on the freezing out of water vapor on the cold surfaces, assumes a value of $\varepsilon=1$ for the cold-side emissivity. In addition, the radiative loading uses simplified geometries consisting of concentric equal-length cylinders and coaxial parallel different-sized disks. It was assumed that the radiation is transferred only between two isolated surfaces, where each surface is gray, diffuse, opaque, and that Kirchhoff's law ($\alpha = \varepsilon$) holds.

The estimated total absolute thermal resistance is 9.2 K/W. This calculation includes the resistance of the various cold fingers, the resistance of the cold plate, and the two copper-on-copper mechanical-joint thermal resistances: between the horizontal and vertical cold fingers (C and H), and between the vertical cold finger and cold plate (H and I). As part of the calculation, the horizontal cold finger is divided into sections depending on changes in cross-sectional area. For example, the cold finger sections containing PEEK stand-offs (D) have less cross-sectional area than the adjoining sections. A conservative conductivity of 1100 W/(m²-K) [13] is used for the joints. For purposes of this calculation, the cold plate, infrared shield, crystal mount, and crystal are considered to be at the same temperature.

## C. Crystal Mount and Wiring Design

The crystal mount is designed to minimize electrical capacitance and insulate the high-voltage surfaces of the crystal from surfaces at ground potential. Specifically, OFHC copper stand-offs (J) distance the cold plate (I) from the high-voltage surfaces and signal read-out contact, and unfilled, virgin PTFE stand-offs (O) insulate the high-voltage crystal contact from the grounded mount. To mechanically support the compliant PTFE stand-offs, OFHC copper backer plates (K) are positioned to put the stand-offs and crystal in compression. As discussed near the end of this section, minimal use of dielectric material close to the crystal is critical for improving noise performance.

The bias voltage is applied to the crystal's outer n+ lithium contact through a thin copper ring located under the PTFE insulator, labelled by the upper O in Fig. 1. Unlike most p-type HPGe detectors, indium is not used in the high-voltage contact in order to eliminate the background produced by the indium's radioactivity. The p+ point contact is instrumented with a commercially available low-noise front-end electronics capsule. This capsule contains an ultra-low noise JFET and optional heater resistor to prevent the transistor from reaching low enough temperatures such that the JFET becomes inoperable. The capsule contains a small pin that makes contact to the crystal's p+ contact patch. A stiff spring is used to apply a constant force between the pin and crystal surface.

Unjacketed miniature coaxial wiring establishes the electrical connection between the front-end electronics capsule and the preamplifier, located outside the vacuum jacket near the dipstick. A total of 6 wires are required to make the electrical connections – substrate, source, drain, feedback, heater resistor, and heater-resistor ground. These wires run the length of the horizontal cold finger and are held firmly in place with small copper plates that are crimped into notches located along the cold finger. The electrical connection between the high-voltage power supply and the copper ring, located on the crystal surface, is made in the same manner except that the coaxial wire's braid has been removed and an additional layer of insulation added in order to achieve the required high-voltage insulation. To prevent the wiring from transferring heat directly to the germanium crystal, two-inch sections of high-thermal-resistance stainless steel wire are added to the leads on the inside of the electrical vacuum feedthrough.

## D. Electrical Considerations

The equivalent noise charge of a charge-sensitive preamplifier is adversely affected by the total capacitance in parallel with its input. Increased capacitance results in poorer energy resolution and thus an increased low-energy threshold [14]. The total capacitance is the sum of intrinsic detector capacitance, preamplifier feedback capacitance, input capacitance of the front-end electronics and any stray capacitance of the interconnections. Modern large-volume PPC-style HPGe detectors can be made with very low intrinsic capacitance (~1 pF) [2], similar to the input capacitance of a high-quality JFET or typical preamplifier feedback capacitance. To take advantage of this low intrinsic capacitance, the cryostat is designed to keep stray capacitance of the interconnections to a small fraction (< 10%) of the other inevitable contributions to the total capacitance.

Fig. 2 and Fig. 3 show electrical potentials calculated with three-dimensional finite-element field-models [15] used to investigate stray capacitance exterior to the germanium crystal. The calculations are performed on a highly detailed model of the crystal mount for a nominal 90-mm diameter, 35-mm tall PPC detector with a 5-mm diameter contact patch. A region of bare germanium on the top crystal surface extends from the contact patch to the inner-edge of the bias-voltage contact.

For Fig. 2, the operating potentials are computed assuming 2500 V bias and a simplistic uniform distribution of p-type donor impurities with a density of $10^{10}$ cm$^{-3}$. The goal is to study stray capacitance exterior to the crystal rather than simulate pulse formation from ionization events in the crystal. The intrinsic capacitance given by this model is 1.8 pF.

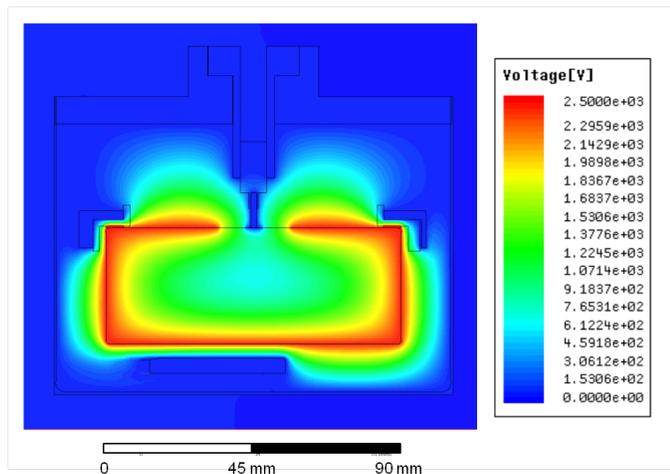

Fig. 2. Electric potentials in plane through axis of crystal produced by 2500 V bias. Potentials are calculated through a finite-element field-model [15].

Fig. 3 shows the electric potentials used in computing the stray capacitance. Following the standard prescription for calculating capacitance, the contact pin and contact patch are held at 1 V with all other conductors, including the bias contact, grounded. The calculated stray capacitance is 0.2 pF. If the input capacitance of a typical JFET (not yet included in the model) is considered, then the stray capacitance of the cryostat satisfies the requirement to be less than 10% of the total.

The total input capacitance affects the preamplifier noise by coupling fluctuating charges from other non-sensitive electrodes according to the usual relation $Q = CV$. However, if any of the capacitance is due to electric fields in dielectrics, then there will be additional *dielectric loss* noise [16]. Therefore significant effort has been made in the models above to include details of the front-end assembly where most of the dielectric loss noise occurs. This source of potential noise is one of the primary reasons the design uses a minimal amount of dielectric material for high-voltage stand-offs. Furthermore, attention is paid to the selection and preparation of the dielectric materials used.

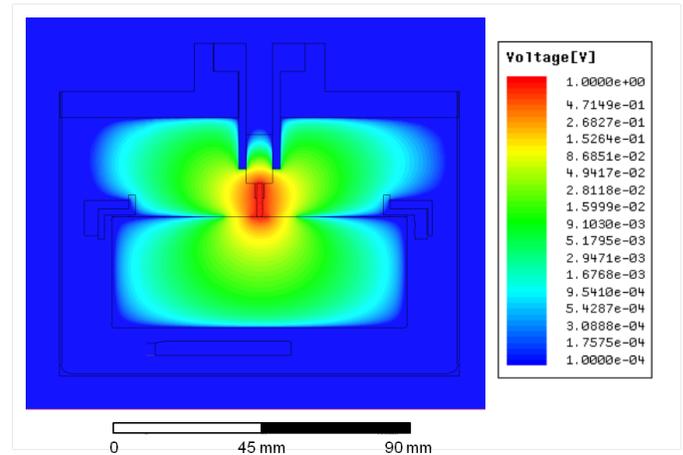

Fig. 3. Electric potentials in plane through axis of crystal used to compute stray capacitance. Potentials are calculated through a finite-element field-model [15]. The capacitance calculation is performed with contact pin and patch held at 1 V with all other conductors at 0 V. To allow for display on a log scale, the image is created with nominally grounded conductors at 0.1 mV.

As part of the cryostat development, Table I reports dimensional stability results for dielectrics used for high-voltage insulation and crystal mounting in germanium detector cryostats when processed with a standardized leaching process [17]. The process includes a 24-hour HN0$_3$ leach, at least a 48-hour vacuum pump, and a 24-hour bake in vacuum.

TABLE I
DIMENSIONAL-STABILITY OF DIELECTRICS

| Dielectric | Process | Geometry | Dimension | Change (0.001") |
|---|---|---|---|---|
| PCTFE | HNO$_3$ | R | OD / ID | 5 / -5 |
| | | C, P | | – |
| | Vacuum | C, P, R | | – |
| | Bake | C | side | -10 |
| | | P | l / t | -20 / 10 |
| | | R | OD / ID | -20 / -15 |
| PTFE | HNO$_3$ | R | OD / ID | 10 / -10 |
| | | C, P | | – |
| | Vacuum | C, P, R | | – |
| | Bake | C, P, R | | – |
| HDPE | HNO$_3$ | C, P, R | | – |
| | Vacuum | R | OD / ID | – / -5 |
| | | C, P | | – |
| | Bake | C | side | 10 |
| | | P | l / t | -10 / 5 |
| | | R | OD / ID | -10 / -5 |
| UHMW | All | C, P, R | | – |
| PEEK | All | C | | – |

For each material and geometry, four individual pieces were tested. Rings (R) had outer diameters (OD) of 76.2 mm (3"),

inner diameters (ID) of 50.4 mm (2"), and thicknesses of 6.35 mm (0.25"). Rectangular plates (P) had side lengths (l) of 50.8 mm (2") and thicknesses (t) of 6.35 mm (0.25"). Cubes (C) had sides of 25.4 mm (1"). Changes were measured in mils (0.001" or 0.0254 mm); no significant change is denoted with a dash (–). Future work will investigate if these dielectrics, in conjunction with the leaching process, negatively impact the electronic noise and low-energy threshold performance of the germanium detector either through dielectric parasitic capacitance or dielectric loss (to ground).

### III. Experimental testing

#### A. Vacuum Testing

In regard to vacuum performance, the cryostat has three areas of concern: the electroformed copper end can, electron-beam welded joints in the crossarm, and demountable o-ring joints. In some cases PNNL electroformed copper has been shown to deposit with micro voids extending through the thickness of the part. To ensure uniform growth and low density of voids the deposition rate is reduced to approximately 50 μm/day. E-beam welding can also be a source of vacuum leaks if the joint is improperly assembled prior to welding. To test the cryostat for leaks a forward-flow leak detector was used with helium tracer gas.

A baseline leak rate of $2.6 \times 10^{-10}$ std. cc/sec was established for the vacuum jacket and dipstick assembly. Helium was then cautiously projected across the surfaces of the cryostat to avoid loading the o-rings with helium. The detector did not respond to the helium, indicating that the vacuum jacket was free of fine leaks. Next, the copper portions of the cryostat were covered with a helium-filled bag until there were indications of o-ring permeation by a slow rise in the leak rate (over a 30-minute period). When the bag was removed and the cryostat blown off with room air, the leak rate slowly decreased, indicating that the previous permeation was through the o-rings.

To test the baseline warm operating pressure, the cryostat was connected to a 67 L/s turbo molecular pump. After one week of pumping and regeneration of the molecular sieve, the cryostat was operating at $3.9 \times 10^{-7}$ torr.

#### B. Thermal Testing

To support comparison to the cryostat thermal model, a series of tests were performed that focused on investigating interface resistance to thermal flow and determining the operating temperature at the end of the dipstick's cold finger. Characterizing this temperature is necessary to understand the performance of the entire cryostat system when assembled. For each test, a solid cylindrical copper test fixture at the end of the cold finger configuration was instrumented with a zener diode and a cryogenic-rated silicon diode. The latter monitored temperature while the former was used to provide a variable heat load. In addition, the mass of the liquid nitrogen used to cool the system was monitored with a platform scale and the measured change used to quantify the power input to the liquid-nitrogen bath. The electrical wires between the zener diode and vacuum feedthrough were each fitted with a 25.4 mm (1") section of high-thermal-resistance stainless steel wire to provide a thermal break while the temperature-sensor wires are made of 36 AWG phosphor-bronze, another high-thermal-resistance material.

In the first set of tests, the test fixture had a diameter of 27.9 mm (1.1") and length of 54.0 mm (2.125"). This fixture was attached to a vendor-supplied horizontal cold finger with a diameter of 19.1 mm (0.75") and length of 168.5 mm (6.635") that in turn was connected to a vertical cold finger within the dipstick with a diameter of 19.1 mm (0.75") and length of 593.7 mm (23.375"). The test fixture was bolted to the cold finger until a Belleville washer, under its head, was flat (125 in-lbs unlubricated). This washer was included to maintain good thermal contact between the surfaces as the temperature cycled.

For each set of tests, four different heat loads were provided through the joule heating of the zener diode with a voltage supplied by a current-limited power supply. The results from the first set of tests are shown in Fig. 4 which plots the measured steady-state temperature vs. zener heat load estimated through two different methods. The solid-square points calculate the load from the rate of change of liquid nitrogen while the open-square points use the approximate P=IV (this equation does not account for heat loss outside the cryostat) where I and V are measured at the power supply. For the liquid-nitrogen boil-off method, data includes subtraction of the base boil-off rate. As can be seen in Fig. 4, the boil-off method indicates loads that are consistently lower by around 14%. For the rest of this paper, only the boil-off method will be employed. The temperature measured at the test fixture with no applied zener power was 82.4 K.

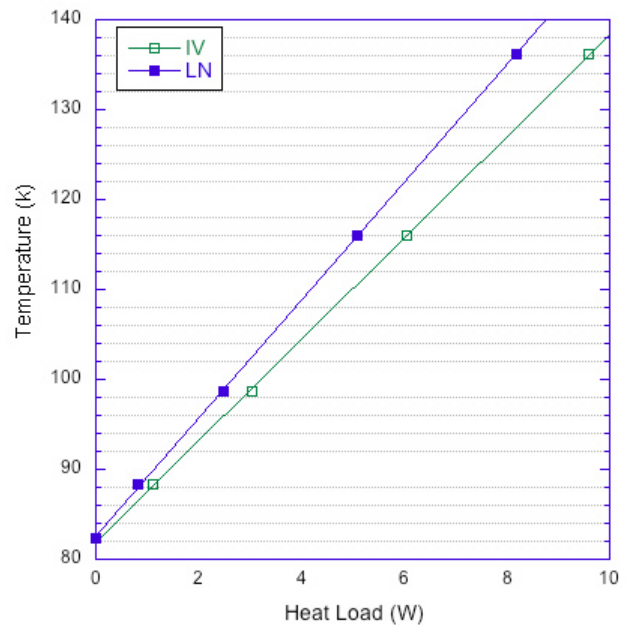

Fig. 4. Plot of temperature vs. heat load for first set of tests (see text). Open-square points estimate heat load with P=IV while solid-square points estimate load from the rate of change of liquid nitrogen. Lines are fits to the data.

The configuration for the second set of tests consisted of the previous set-up along with a solid cylindrical copper extension with a diameter of 27.9 mm (1.1") and length of 25.4 mm (1") inserted between the horizontal cold finger and test fixture. A bolt with a Belleville washer connected the test fixture and extension to the horizontal cold finger. Fig. 5 displays the plot of the temperature vs. heat load (estimated through boil-off) for the first two sets of tests, i.e., with and without the extension. The base temperature increased by 3.7 K and the slope was essentially unchanged.

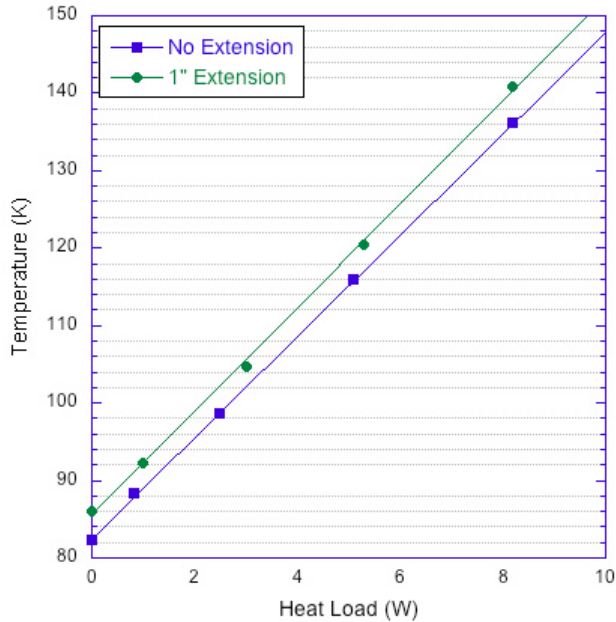

Fig. 5. Plot of temperature vs. heat load (estimated through boil-off) for first two sets of tests (see text). Lines are fits to the data. A 3.7 K temperature shift is seen between the tests.

For the third set of tests, a 75 µm-thick indium layer was inserted between the test fixture and extension before tightening the joint. Fig. 6 includes the results from this set of tests whose purpose was to check whether or not indium improved the thermal conductance of the joint. The fact that the data from the indium test lies on top of the data from the "no extension" test indicates that the joint containing indium is essentially thermally non-existent. The main difference between these two configurations is an indium-filled joint.

Next, the three elements comprising the previous horizontal cold-finger section were replaced by a single solid cylindrical test fixture with diameter 27.9 mm (1.1") and length 235.0 mm (9.25"). This fixture was instrumented with the same diodes as before and the diameter chosen to match the design diameter of the cryostat's horizontal cold finger. A 75 µm-thick indium layer was inserted in the joint between the dipstick's vertical cold finger and the new test fixture (the original joint with the vertical cold finger also had an indium layer supplied by the vendor, though the thickness is unknown). This is the first configuration to have only a single thermal joint. Once again, a bolt with a Belleville washer secured the joint.

Fig. 7 displays the plot of temperature vs. heat load for all four sets of tests. The linear (least-squares) fit to the data from the last test has a reduced slope as compared to the other tests, which is to be expected from the last configuration's lower thermal resistance. We choose to use these measurements to discuss the uncertainties present in the data.

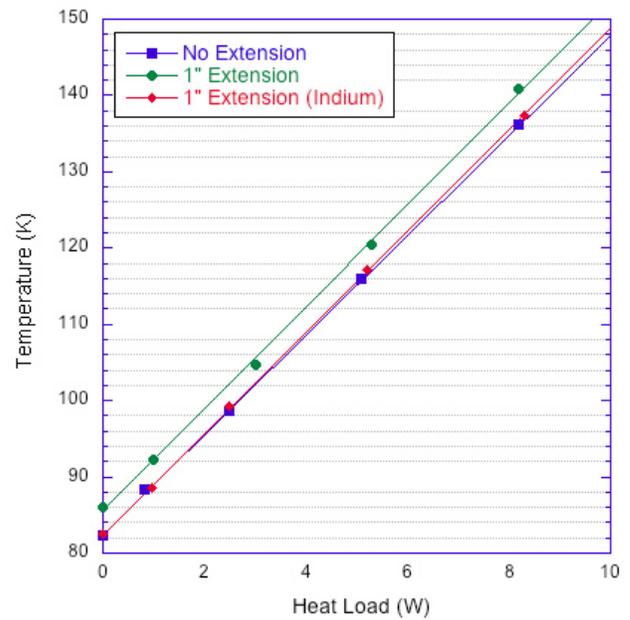

Fig. 6. Plot of temperature vs. heat load for first three sets of tests (see text).

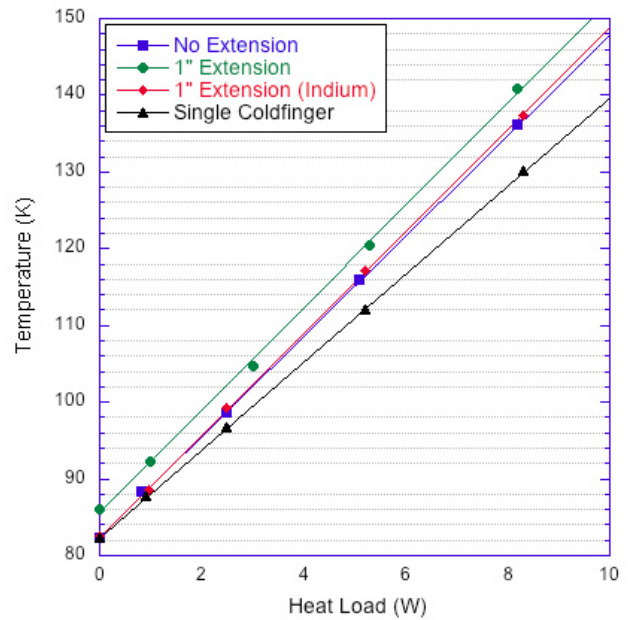

Fig. 7. Plot of temperature vs. heat load for first four sets of tests (see text).

The temperature (ordinate) errors consist of slow drifts (likely due to changes in the ambient temperature) and systematic uncertainties intrinsic to the temperature sensor. The slow drifts provide an error of at most 0.1 K over the timescales involved and the sensor documentation specifies the sensor systematic error as 0.2 K. The latter error has negligible influence on the slope as the systematic uncertainty changes each temperature reading by roughly the same amount. The heat-load (abscissa) error is determined by utilizing the standard uncertainty in the slope parameter from the least-squares fit to the liquid-nitrogen mass vs. time plot. This gives fractional errors in the heat load of around 0.5%.

Incorporating these errors into a worst-case estimate for the uncertainty in the slope of the last test's temperature vs. heat-load line, gives a 1.2% fractional error. Specifically, the "1 watt" and "9 watt" points are shifted in both abscissa and ordinate by the appropriate uncertainty so as to maximize the change in slope. If the error in the slope of each line is estimated instead by utilizing the scatter (around a linear fit) in the temperature vs. heat-load points, then fractional errors of 0.8%, 1.6%, 0.6%, and 0.3% are obtained respectively for the four sets of tests. Uncertainties at this level (~1%) are too small to be seen on the plots.

Results from the thermal data were compared to predictions from thermal modeling similar to that presented for the cryostat design. The only data that does not have an adequate explanation is the base-temperature shift measured between test 1 and test 2. The magnitude of this shift is inconsistent with the lack of a change in slope. Ignoring this for the present, the rest of these data imply three conclusions regarding the thermal resistances of the mechanical joints. First, as mentioned before, addition of a 75-μm thick indium layer essentially eliminates the thermal resistance. Second, the thermal resistance of a copper-on-copper joint is not reproducible, i.e. changes between assemblies. This result is consistent with other studies [13]. Third, the various copper-on-copper joints had thermal resistances between 0.5 K/W and 0.9 K/W. Magnitudes such as these are a principal reason the horizontal cold finger is designed to be a single piece. Finally, it is clear future efforts to fully understand the thermal modeling of mechanical joints in cryogenic systems will require instrumentation (temperature and heat load) on each and every portion of the primary thermal cooling path.

## IV. CONCLUSION

Progress has been made on a cryostat intended for ultra-low-energy threshold p-type point contact germanium spectrometers. The vacuum envelope has been assembled, tested, and found to meet design requirements. Initial thermal tests have investigated the operating temperature at the end of the dipstick's cold finger as well as mechanical-joint thermal resistances. The base dipstick temperature has been measured to be 82.4 K. It was confirmed that an indium layer significantly reduces a joint's thermal resistance and a range of magnitudes was determined for the thermal resistances of the copper-on-copper joints. These ongoing thermal tests will eventually allow a quantitative comparison to the presented thermal model.

Next steps include full assembly and testing of the thermal system. The cryostat will then be ready for insertion of different crystal mounts and/or crystals to investigate how low of an energy threshold we can achieve. The presented mount design and electric-potential models suggest a significant improvement is possible.

## ACKNOWLEDGMENTS

We thank J. I. Collar for many helpful discussions on the issues facing energy threshold reduction in large volume p-type point contact germanium detectors. We greatly appreciate input and comments from K. M. Yocum and J. Colaresi on our cryostat design, especially related to making some features of the cryostat amenable to using commercially available mounts, front-ends, and liquid-nitrogen dipsticks. Thanks also to E. W. Hoppe and J. Merriman for fabrication of the cryostat end cap and infrared shield using their copper electroforming capability.